# The flux pinning force and vortex phase diagram of single crystal FeTe$_{0.60}$Se$_{0.40}$

C. S. Yadav and P. L. Paulose

*Department of Condensed Matter Physics & Material Sciences, Tata Institute of Fundamental Research, Colaba, Mumbai-400005, INDIA*

E mail: csyadav@tifr.res.in, paulose@tifr.res.in

**Abstract**

The flux pinning force density ($F_p$) of the single crystalline FeTe$_{0.60}$Se$_{0.40}$ superconductor has been calculated from the magnetization measurements. The normalized $F_p$ versus $h$ ($=H/H_{irr}$) curves are scaled using the Dew-Hughes's formula $f(h) \sim h^p(1-h)^q$ to underline the pinning mechanism in the compound. The obtained values of pinning parameters 'p' and 'q' indicate the vortex pinning by the mixing of the surface and the point core pinning of the normal centers. The vortex phase diagram has also been drawn for the first time for the FeTe$_{0.60}$Se$_{0.40}$, which has very high values of critical current density $J_c \sim 1\times10^5$ Amp/cm$^2$ and the upper critical field H$_{c2}$(0) $\sim$ 65T, with a reasonably high transition temperature T$_c$ =14.5K.

Key words: Superconductors, flux pinning, pinning force, vortex phase diagram

The discovery of superconductivity at 26K in the Iron based pnictide LaFeAsO$_{1-x}$F$_x$ (1111) has reignited the research interests in the field of superconductivity.[1] The substitution of La by other lanthanides (Ce, Nd, Sn) and fluorine free Gd$_{1-x}$Th$_x$FeAs compound raised the transition temperature to 56K, which is the highest reported till date for the non-cuprate superconductors.[2] This also lead to the discovery of superconductivity in other compounds viz. (BaSr)$_{1-x}$K$_x$Fe$_2$As$_2$ (122), the ternary LiFeAs (111) and the binary FeSe/Te (11).[3-6] The Fe based materials show good metallic properties compared to the cuprate superconductors. All these compounds have tetragonal structure with either FeAs or FeTe/Se layers that play a pivotal role in the superconductivity of these materials, similar to the role of Cu-O plane in high temperature superconductors. It would be important from potential application point of view to study the vortex phase diagram and explore its relation to that of high Tc cuprates and other superconductors.

Among the Fe based superconductors, the compound FeSe/Te has been very promising from the point of understanding the real mechanism of superconductivity in this class of compound and also for the technological applications. The FeSe shows the transition temperature $T_c$ at 8K, which rises to to 14.5K for 60% substitution of Se by Te.[7] The nuclear magnetic resonance studies (NMR) revealed FeSe to be the first binary compound to show unconventional superconductivity in the d-electron systems.[8] The penetration depth measurement performed using the muon spin resonance (µSR) technique also favors the presence of the two gaps and anisotropic order parameter symmetry.[9]

In the present work, we have studied the pinning mechanism of the flux density and have drawn the vortex phase diagram for the FeTe$_{0.60}$Se$_{0.40}$ single crystal, which has the highest Tc (~14.5K) among the FeTe/Se superconductors. The high values of upper critical field (H$_{c2}$(0) ~

65T) and the critical current density ($J_c \sim 1\times10^5$ Amp/cm$^2$) makes it suitable candidate for the technological applications.[2,10]

The single crystals of FeTe$_{0.60}$Se$_{0.40}$ compound were prepared by the chemical reaction of the high purity elements (Fe chunk of 99.999% purity, Te powder of 99.99% purity and Se powder of 99.98% purity) in the stochiometric proportion, inside the evacuated quartz tube. The charge was slowly heated to 950$^0$C at the rate 50$^0$C/hrs and kept for 12 hours before cooling down to 400$^0$C at the rate of 6$^0$C/hrs, and then furnace cooled to the room temperature. The obtained crystals were found to grow along the ab-basal plane. The detailed sample preparation method and the quality of the crystal are reported elsewhere.[10] Magnetization measurements have been performed on the high quality single crystal of FeTe$_{0.60}$Se$_{0.40}$. The isothermal Magnetic hysteresis loops (MHL) were obtained using the Vibration Sample Magnetometer (VSM) from OXFORD instruments, up to 12 Tesla magnetic fields.

The MHL curves for FeTe$_{0.60}$Se$_{0.40}$ show second magnetization peak (known as fishtail effect) for both directions along the ab-plane and c axis of crystal.[10] The critical current densities $J_c(T,H)$, can be calculated from these MHL isotherms, using the Bean's critical state model, with the formula $J_c = 20.\Delta M /\{a(1- a/3b)\}$, where 'a' and' b' are the sample dimensions perpendicular to the field direction, and $\Delta M$ is the difference between the magnetization measured while decreasing and increasing magnetic field.[11] The $J_c(T,H)$ for the field ($H$) parallel to c axis ($H//c$) of the crystal are shown in the figure 1(a). Since the value of the value of $J_c(T,H)$ do not vary much with the magnetic field for low temperature isotherms, the Bean's model which is applicable for field independent $J_c$ gives a good and comparative estimation of the critical current density.

The pinning force density ($F_p$) of the compound is calculated for the H//c directions. The pinning force per unit length of the pinning flux line is defined as the work done in moving the flux line from a pinning center to the unpinned position, and is given as $F_p \propto L\Delta W/x$; where $L$ is the length of the flux line, $\Delta W$ is the work done and $x$ is the range of pinning interaction. Experimentally, the pinning force density $F_p$ is calculated from the $J_c(T,H)$ by using the formula $F_p = J_c \times \mu_0 H$. We have shown the field dependence of $J_c(T,H)$ and $F_p$ for *H//c* direction in the figure 1. In the measured field range up to 12T, the pinning force shows a maxima at all temperatures (2<T<11K). To elucidate the mechanism of the pinning force, Dew-Hughes proposed a scaling rule $f_p \propto h^p(1-h)^q$ for the conventional BCS type superconductors.[12] Here $f_p$ and $h$ are the normalized pinning force and the field, and are defined as $f_p = F_p/F_{p,max}$ and $h = H/H_{c2}$. Similar scaling was also found to valid in the case of unconventional high temperature superconductors (HTSC), with $h = H/H_{irr}$, where $H_{irr}$ is the irreversibility field.[13-15] The $H_{irr}$ is taken as the value of applied magnetic field, where $J_c(T,H)$ becomes too small to be useful for any practical application.

The scaled plots of $f_p$ versus $h$ for 6<T<10K are shown in the figure 2. We have used the criterion of $J_c < 50 Amp/cm^2$ for determining $H_{irr}$. As seen from the figure 2, $f_p$ curves for different temperatures do not fall together at high magnetic field. The dotted line in the figure shows the best possible fitted curve for $f_p = A h^p (1 - h)^q$ with A = 27.9, $p$ =1.54 and $q$ =3.8. The value p/(p+q) ~0.28, matches well with the peak positions of these $f_p$ versus $h$ plots. These values of exponent $p$ and $q$ do not fit into any single model for flux pinning mechanism by Dew-Hughes.[12] However, similar values of $p$ and $q$ has been reported for other unconventional superconductors viz. $NdBa_2Cu_3O_{7-\delta}$ (p=1.48, q=2.23, $h_{max}$=0.45); $YBa_2Cu_3O_7$ (p=2, q=4, $h_{max}$=0.33), and $Sm_{0.5}Eu_{0.5}Ba_2Cu_3O_{7-\delta}$ (p=2.08, q=3.56, $h_{max}$ =0.37).[14-16] The value of $h_{max}$ ~0.28, for our

FeTe$_{0.60}$Se$_{0.40}$ single crystal. Dew Hughes has propounded a model where $h_{max}$=0.2 and 0.33 correspond to the surface pinning and the point core pinning respectively.[12] Thus $h_{max}$~0.28 can be understood in terms of $\delta l$ type pinning with a mixture of the surface and the point core pinning of the normal centers, with the different range of the pinning interactions. There is a clear deviation of the f$_p$ versus h curves from the fitted plot at higher magnetic fields which indicates towards the onset of flux creep phenomenon with the occurrence of the small size normal pinning.[17] In the inset of figure 2, we have plotted F$_{max}$ versus H$_{max}$, showing $F_{max}$=0.45 $H_{max}^2$ kind of dependence. Therefore the pinning force density can be scaled as $F_p = 0.45\ H_{max}^2 \cdot f_p$.

Vortex phase diagram for the compound is shown in the figure 3, where we have shown it for three characteristic fields H$_{min}$, H$_{sp}$ and H$_{irr}$. The H$_{min}$, H$_{sp}$ are the magnetic fields located at the valley point and the second magnetization peak (sp) respectively, in the $J_c$ versus H plot, as also explained in the inset of the figure 3. The $H_{min}$-T, $H_{sp}$-T and $H_{irr}$-T curves can be fitted using the empirical expression $H(T)=H(0).(1-T/T_c)^n$. the parameter obtained are H$_{min}$(0)=1.45T, H$_{sp}$(0)=12.17T, and H$_{irr}$(0) = 87.51T; and n = 2.1, 3.02 and 3.06 for $H_{min}$-T , $H_{sp}$-T and $H_{irr}$-T curves respectively. The lower critical field (H$_{c1}$) of the FeTe$_{0.60}$Se$_{0.40}$ crystal measured along the H//c and H//ab (shown in figure 4) are also found to have high values of power law exponent n ~ 2.46 and n ~ 1.99. The detailed procedure for H$_{c1}$ determination is reported elsewhere.[10] The vortex phase diagram of the compound shows a very high value of the exponent 'n' which cannot be understood within the framework of the conventional superconductors. These high values of exponent 'n' for the FeTe$_{0.60}$Se$_{0.40}$ found their analogue in Tl based cuprates superconductors were 'n' is reported to be 3.4 and 2.5 for the TlCa$_3$Ba$_2$Cu$_4$O$_{11-\delta}$ and (TlBi)1223 thin film respectively.[18, 19]

We have calculated the flux pinning force density and have shown the existence of surface and point core pinning of the normal centers in the low field regime, and the occurrence the flux creep in high field regime. The vortex phase diagram has been drawn for the compound. The high value of the exponent 'n' in the $H(T) \sim (1-T/T_c)^n$ indicate towards the non-conventional nature of the superconductivity in this compound.

**FIGURE CAPTIONS**

FIG 1. Magnetic field dependence of the (a) Critical current density $J_c$ (as semi-logarithmic plot), and (b) pinning force density $F_p=J_c \times H$ measured in FeTe$_{0.60}$Se$_{0.40}$ single crystal at various temperatures in fields parallel to the c axis.

FIG 2. The scaled density of the pinning force $f_p$ (=$F_p/F_{p\ max}$) as a function of the reduced field, h = H /$H_{c2}$. Dotted line represents the fitted curve $h^{1.54}(1-h)^{3.8}$. Inset: $F_{max}$ as a function of $H_{max}$. The solid line in the inset is a fit of $F_{max} = 0.45\ H_{max}^2$.

FIG 3. The vortex phase diagram of the FeTe$_{0.60}$Se$_{0.40}$ single crystal for H//c. The dotted lines show the fit to curve $H(T)=H_0(1-T/T_c)^n$. Inset of the figure explains the point taken for the $H_{irr}$, $H_{sp}$ and $H_{min}$ values on the $J_c$ vs H curve.

FIG 4. The lower critical field ($H_{c1}$) versus T curve for FeTe$_{0.60}$Se$_{0.40}$ single crystal for H//c and H//ab direction.

**Figure 1.**

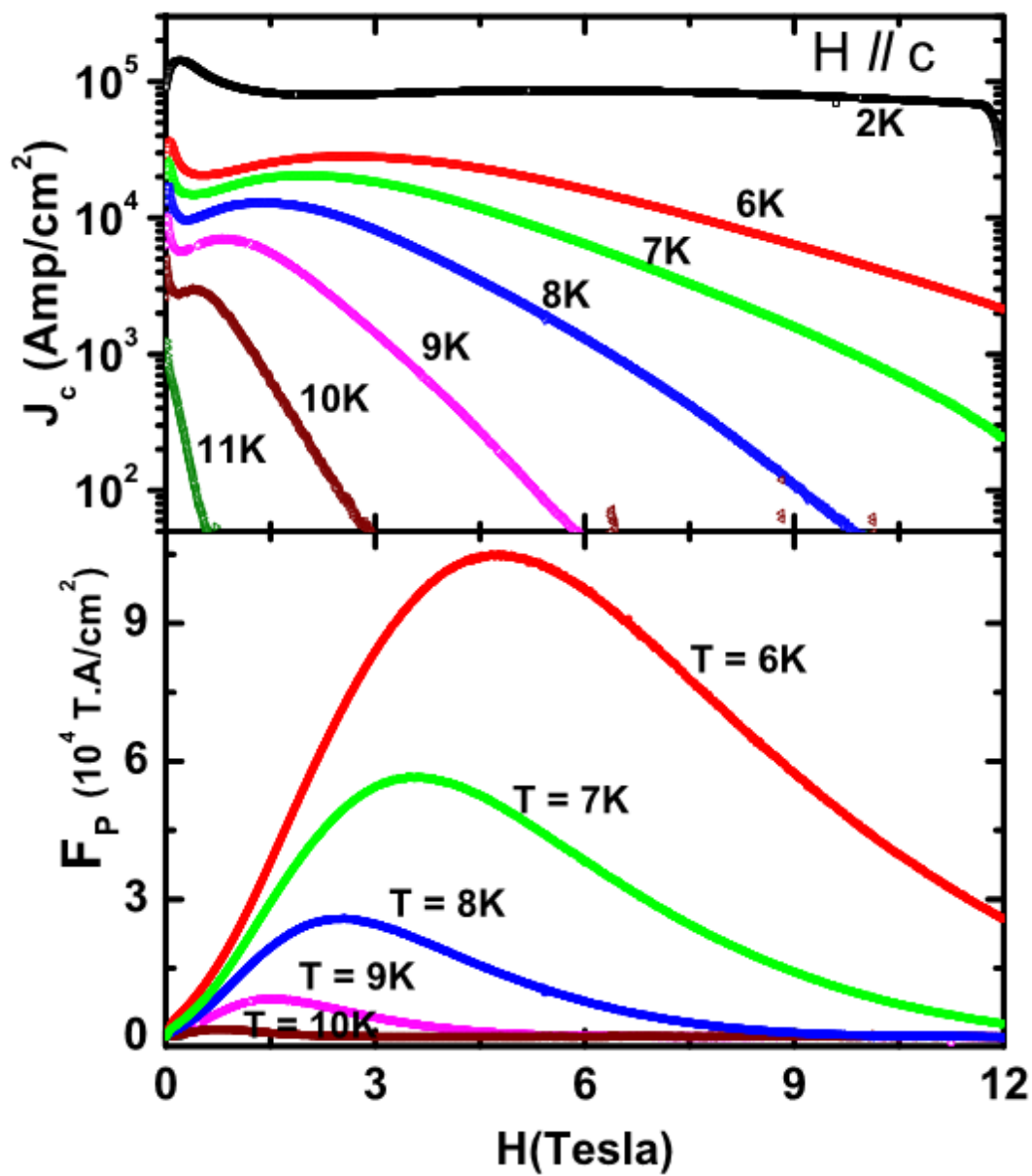

**Figure 2.**

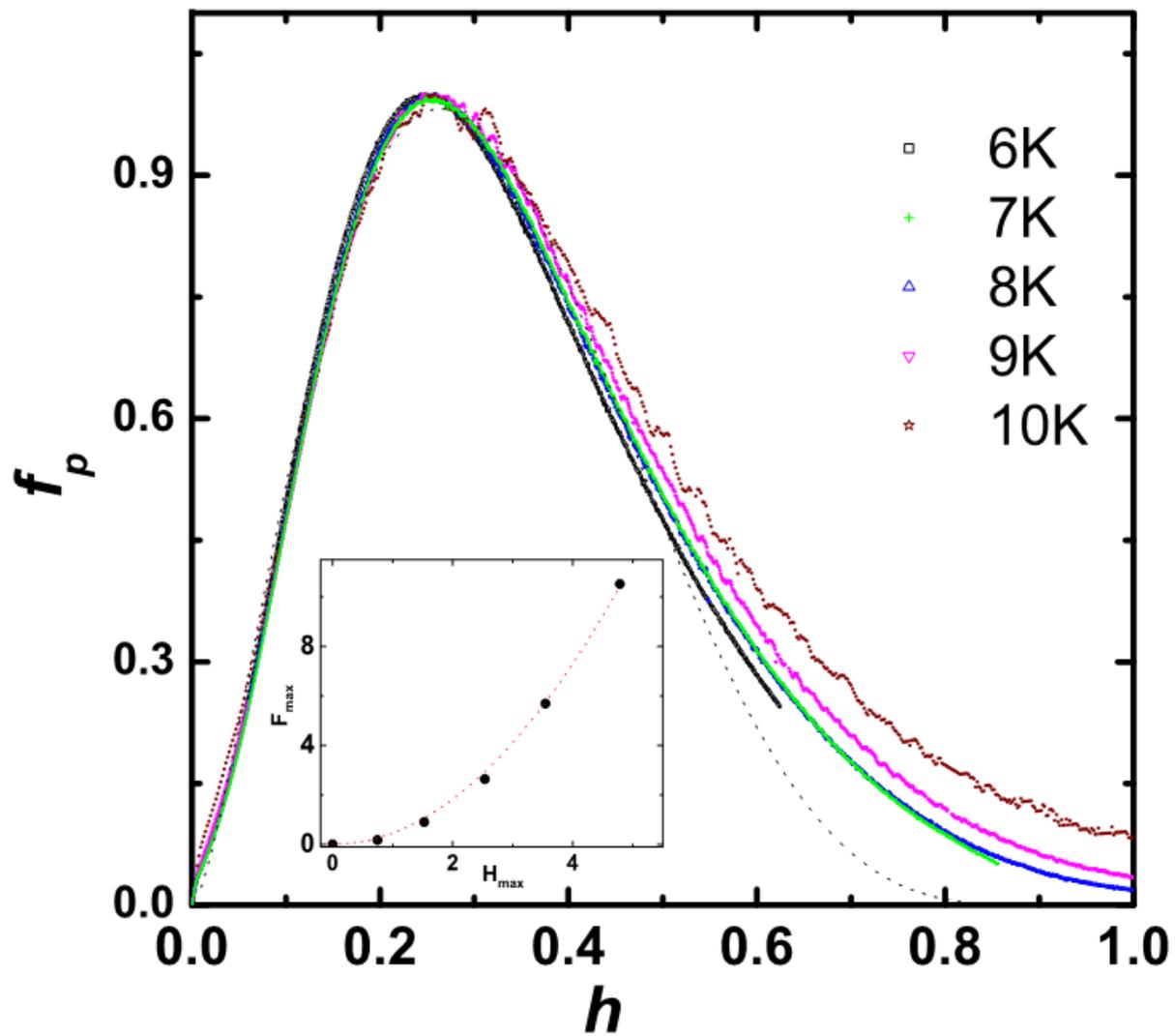

**Figure 3.**

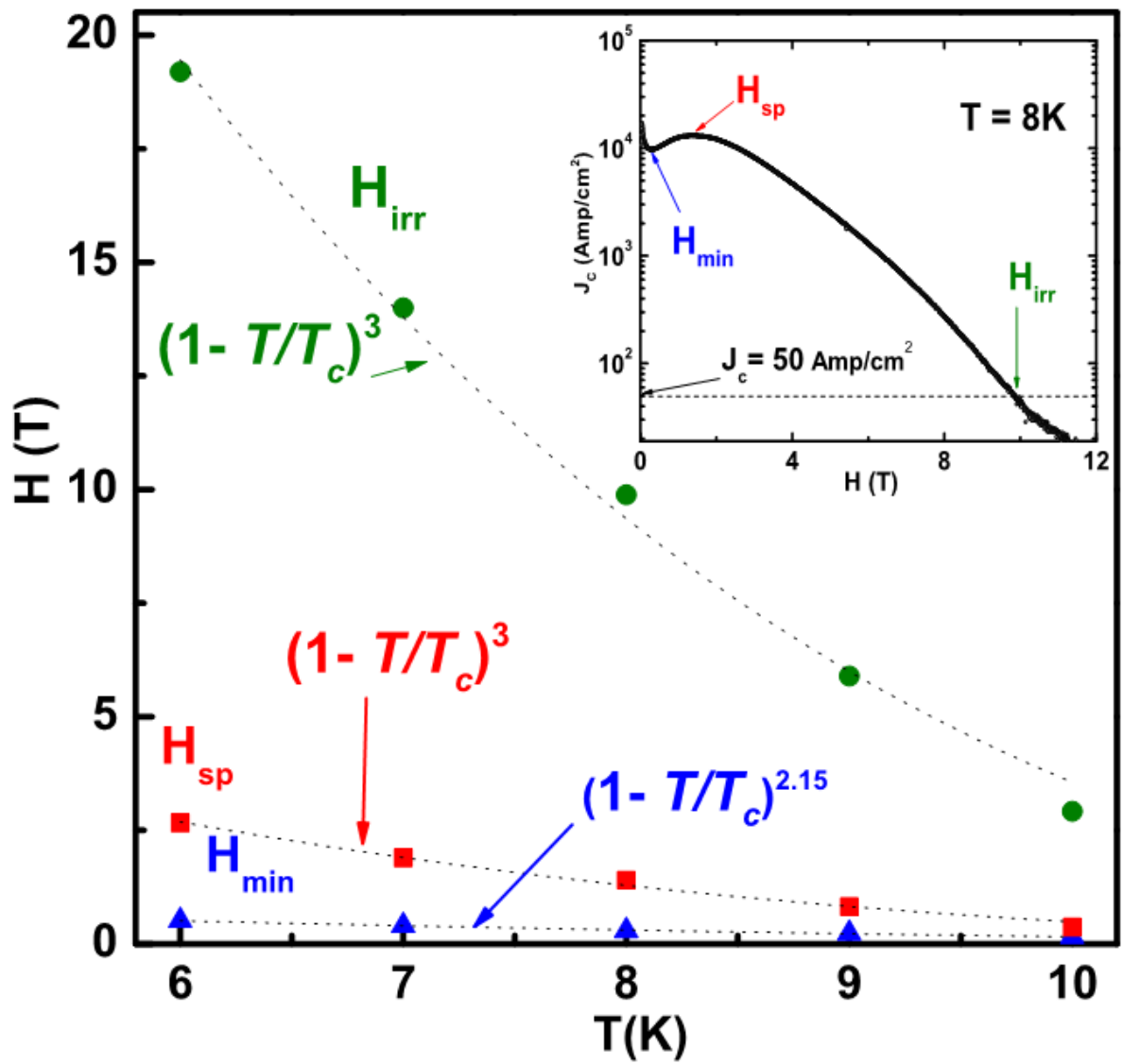

**Figure 4.**

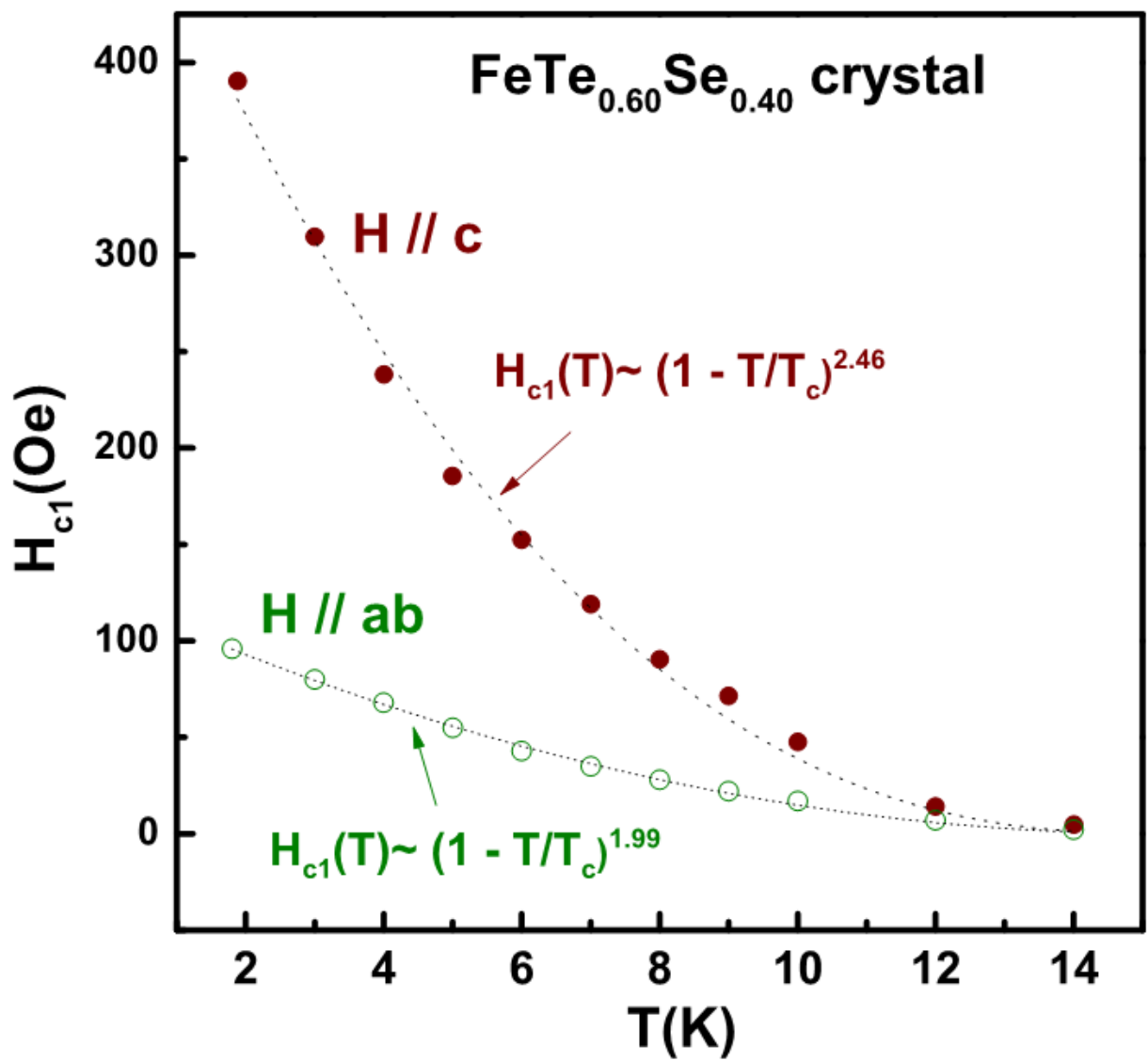